\documentclass[traditabstract]{aa} 
\usepackage{graphicx}
\usepackage{txfonts}
\usepackage{natbib}
\usepackage{color}
\begin{document}

   \title{The subgiant branch of $\omega$~Cen seen through high-resolution
	  spectroscopy. II. The most metal-rich population\thanks{Based on
	  observations with the ESO GIRAFFE@VLT, under programme 079.D-0021(A).
	  Also based on literature data obtained with WFI@VLT (programmes 
	  62.L-0345 and 63.L-0439) and FORS@VLT (programme 68.D-0332).  The
	  following online databases were also extensively employed:  NIST,
	  VALD, Kurucz.}\fnmsep\thanks{Full Table~\ref{tab-ew} is only
	  available in electronic form at the CDS via anonymous ftp to
	  cdsarc.u-strasbg.fr (130.79.128.5) or via
	  http://cdsarc.u-strasbg.fr/viz-bin/qcat?J/A+A/XXX/XXX}}

\titlerunning{The most metal-rich population in $\omega$~Centauri} 

   \author{E. Pancino \inst{1},  A. Mucciarelli\inst{2}, 
           P. Bonifacio\inst{3}, L. Monaco\inst{4}, and L. Sbordone\inst{3,5}}
   \authorrunning{Pancino et al.}
   \institute{INAF-Osservatorio Astronomico di Bologna, via Ranzani 1, I-40127,
	     Bologna, Italy\\
             \email{elena.pancino@oabo.inaf.it}
             \and
	Dipartimento di Astronomia, Universit\`a degli studi di Bologna, via
	     Ranzani 1, I-40127, Bologna, Italy 
	     \and
	GEPI, Observatoire de Paris, CNRS, Univ. Paris Diderot, 5 Place Jules
             Janssen, F-92195 Meudon, France
	     \and
	European Southern Observatory, Casilla 19100, Santiago, Chile  
	     \and     
	Max Planck Institute for Astrophysics, Karl-Schwarzschild-Str. 1, D-85741
             Garching, Germany  
             }

   \date{Received September XX, XXXX; accepted March XX, XXXX}

 
   \abstract{
   
   We analyze spectra of 18 stars belonging to the faintest subgiant branch in
   $\omega$~Centauri (the SGB-a), obtained with GIRAFFE@VLT at a resolution of
   R$\simeq$17\,000 and a S/N ratio between 25 and 50. We measure abundances of
   Al, Ba, Ca, Fe, Ni, Si, and Ti and we find that these stars have
   $<$[Fe/H]$>$=--0.73$\pm$0.14~dex, similarly to the corresponding red giant
   branch population (the RGB-a). We also measure
   $<$[$\alpha$/Fe]$>$=+0.40$\pm$0.16~dex, and $<$[Ba/Fe]$>$=+0.87$\pm$0.23~dex,
   in general agreement with past studies. It is very interesting to note that
   we found a uniform Al abundance, $<$[Al/Fe]$>$=+0.32$\pm$0.14~dex, for all
   the 18 SGB-a stars analysed here, thus supporting past evidence that the
   usual (anti-)correlations are not present in this population, and suggesting
   a non globular cluster-like origin of this particular population. In the
   dwarf galaxy hypothesis for the formation of $\omega$~Cen, this population
   might be the best candidate for the field population of its putative parent
   galaxy, although some of its properties appear contradictory. It has also
   been suggested that the most metal-rich population in $\omega$~Cen is
   significantly enriched in helium. If this is true, the traditional abundance
   analysis techniques, based on model atmospheres with normal helium content,
   might lead to errors. We have computed helium enhanced atmospheres for three
   stars in our sample and verified that the abundance errors due to the use of
   non-enhanced atmospheres are negligible. Additional, indirect support to the
   enhanced helium content of the SGB-a population comes from our Li upper
   limits.

   }


   \maketitle
%

\section{Introduction}
\label{sec-intro}

The region around the main sequence (hereafter MS) turn off (TO) and the subgiant
branch (SGB) is the most sensitive of the colour-magnitude diagram (CMD) to age,
and as such it has been the target of several studies to disentangle the
age-metallicity relation in the complex mix of stellar populations of
$\omega$~Centauri. The first studies employed photometric metallicity indicators
\citep{hilker00,hughes00}, soon followed by low-resolution spectroscopic samples
\citep{hughes04,hilker04,rey04,sollima05b,stanford06,villanova07}, coupled with
exquisite new photometric catalogues \citep{ferraro04,bedin04,bellini10}, but the
puzzle got deeper, as extensively discussed in the cited papers, and in the first
paper of this series \citep[][hereafter Paper~I]{p11}.

We focus here on the SGB-a, a sub-structure of the SGB of $\omega$ Cen, first
found by \citet{ferraro04}, and also named branch D by \citet{villanova07}. The
SGB-a appears to have a fainter magnitude than all other SGB components, and
merges into the MS of $\omega$~Cen at a fainter magnitude than the TO of all
other populations. There is nowadays no doubt left that the SGB-a is
photometrically the same population as the RGB-a, the reddest and most
metal-rich component of the red giant branch (RGB) of $\omega$ Cen
\citep{p00,p02}, as clearly visible in most recent high-quality photometries
\citep[see, e.g.][]{bedin04,villanova07,bellini10}. However, there is still some
residual debate about its metallicity. High-resolution spectroscopic studies of
RGB stars always find a metal-rich component with [Fe/H] higher than --1.0~dex,
and generally around --0.6 or --0.7~dex on average
\citep{p02,p03,p04,johnson10,marino11}. Only two studies were performed on SGB-a
stars, both based on low-resolution spectroscopy (R$\simeq$6000), and they give
quite discrepant results. The former \citep{sollima05b}, based on calcium
triplet measurements, found [Fe/H]$\simeq$--0.6~dex, while the latter
\citep{villanova07}  found [Fe/H]$\simeq$--1.1~dex, with no star with [Fe/H]
above --1.0~dex in their sample, even if a handful of their targets belonged to
the SGB-a, or branch D\footnote{A possible call to caution in this respect
comes from  \citet{bellini10}, who found that the SGB-a could be split into two
very close branches (see their Figures~7, 9, 10, and 11). Therefore, the
properties of this population could be more complicated than expected.}.

\begin{table*}
\caption{Observing Logs and atmospheric parameters.}
\label{tab_logs}
\begin{center}
\begin{tabular}{l c c c c c c c c c c c c} 
\hline\hline
\noalign{\smallskip}
ID$_{WFI}^a$ & ID$_{FORS}^b$ & R.A. (J2000) & Dec (J2000) & V$^b$ & (B--V)$_0^b$ & 
     (V--I$_J$)$_0^b$ & $n_{obs}$ & S/N & T$_{eff}^{H_{\alpha}}$ & log$g$ & v$_t$  \\
& & (deg) & (deg) & (mag) & (mag) & (mag) & & & (K) & (dex) & (km s$^{-1}$) \\
\noalign{\smallskip}
\hline
\noalign{\smallskip}
213129 & 33807 & 201.7662757 & -47.5488277 & 17.73 & 0.70 & 0.99 & 1 & 35 & 5300 & 3.8 & 1.0 \\
213295 & 40145 & 201.6777863 & -47.5484132 & 17.98 & 0.66 & 0.91 & 9 & 45 & 5500 & 3.9 & 1.0 \\
214198 & 40987 & 201.6423287 & -47.5449255 & 18.22 & 0.62 & 0.88 & 1 & 26 & 5450 & 4.0 & 1.0 \\
215315 & 65607 & 201.7166701 & -47.5407130 & 17.75 & 0.68 & 0.85 & 1 & 35 & 5650 & 3.9 & 1.0 \\
215700 & 37198 & 201.6256851 & -47.5393905 & 18.11 & 0.62 & 0.89 & 9 & 49 & 5500 & 4.0 & 1.0 \\
215931 & 61617 & 201.7105924 & -47.5383614 & 18.05 & 0.67 & 0.86 & 1 & 37 & 5650 & 4.0 & 1.0 \\
216031 & 34467 & 201.6134380 & -47.5381179 & 17.65 & 0.71 & 1.03 & 9 & 49 & 5250 & 3.7 & 1.0 \\
218364 & 29483 & 201.7772506 & -47.5289457 & 17.50 & 0.74 & 0.98 & 1 & 33 & 5300 & 3.7 & 1.0 \\
220401 & 42942 & 201.6264389 & -47.5217612 & 18.22 & 0.64 & 0.99 & 1 & 25 & 5250 & 3.9 & 1.0 \\
220947 & 29233 & 201.7634185 & -47.5195723 & 17.57 & 0.77 & 1.03 & 1 & 37 & 5500 & 3.8 & 1.0 \\
224701 & 63733 & 201.7160750 & -47.5054148 & 17.86 & 0.73 & 0.89 & 1 & 38 & 5500 & 3.9 & 1.0 \\
224921 & 33448 & 201.7852963 & -47.5044375 & 17.98 & 0.65 & 0.96 & 9 & 47 & 5500 & 3.9 & 1.0 \\
227902 & 30356 & 201.7883906 & -47.4929578 & 18.21 & 0.65 & 0.61 & 1 & 32 & 5750 & 4.1 & 1.0 \\
234254 & 33737 & 201.6024956 & -47.4688537 & 17.97 & 0.64 & 0.99 & 1 & 30 & 5350 & 3.9 & 1.0 \\
235569 & 62919 & 201.6354387 & -47.4637003 & 17.95 & 0.69 & 0.98 & 1 & 39 & 5650 & 4.0 & 1.0 \\
243327 & 37860 & 201.6368456 & -47.4344181 & 17.63 & 0.73 & 1.04 & 9 & 48 & 5300 & 3.7 & 1.0 \\
247798 & 37337 & 201.6311023 & -47.4175838 & 18.24 & 0.60 & 0.96 & 9 & 45 & 5500 & 4.0 & 1.0 \\
248814 & 38329 & 201.7716525 & -47.4135293 & 18.08 & 0.62 & 0.89 & 9 & 43 & 5500 & 4.0 & 1.0 \\
\hline\hline	                                                                                                                      
\multicolumn{9}{l}{$^a$Photometry from \citet{p00}.}\\                                         
\multicolumn{9}{l}{$^b$Photometry from \citet{sollima05a}.}\\                                  
\end{tabular}                                                                                  
\end{center}                                                                                   
\end{table*}                                                                                   

\begin{figure}
\centering
\includegraphics[scale=0.3]{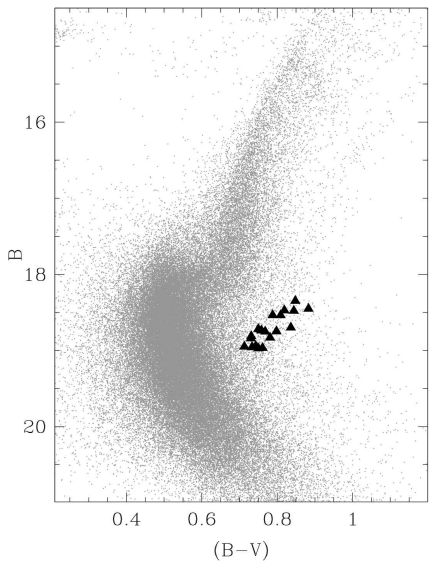}
\caption{Selected SGB-a stars (black triangles) on the FORS
photometry (grey dots) by \citet{sollima05a}.}
\label{fig_cmds}
\end{figure}

Besides the problem of the SGB-a metallicity, there is also some doubt about the
$\alpha$-elements enhancement of this metal-rich population. Two studies found
that RGB-a stars had lower [$\alpha$/Fe] with respect to MInt and MP stars
\citep[metal-intermediate and metal-poor, respectively, following the definition
by][]{p00}, namely \citet{p02} and \citet{origlia03}, the former containing three
RGB-a stars studied with UVES and the latter more stars, but with SOFI
low-resolution infrared spectroscopy. The confirmation of  such a finding would
be extremely important, because it would imply -- in the self-enrichment
framework for the formation of $\omega$~Cen -- that type Ia supernovae had a
chance to contribute to the chemical enrichment of this population.
Unfortunately, more recent results cast some doubt on this result, or at least on
its quantitative evaluation. In particular, \citet{johnson10} found that a few
stars in common with \citet{p02} had a higher [$\alpha$/Fe] than previously
thought. It is clear, however, that a peculiar trend is visible in the behaviour
of a few $\alpha$-elements, and in particular in [Ca/Fe] \citep[see
e.g.,][]{norris95,smith04,p04}, where the enhancement is higher for RGB-MInt
stars than for RGB-MP and RGB-a stars. In other words, [Ca/Fe] rises slightly
with metallicity, reaches its maximum value around [Fe/H]$\simeq$--1.2~dex, and
then gently decreases again. The amount of this variation is small, around
0.1~dex, and certainly below 0.2~dex, but it is clearly visible in the
high-resolution study by \citet{johnson10}, as well. Such a trend, if confirmed,
is not straightforward to interpret -- having most probably to do with a variable
star formation rate -- and it requires a full modelization of the chemical
evolution of $\omega$~Cen. In fact, while the intervention of type Ia supernovae
could still be the best way to explain the (slight) decrease in [$\alpha$/Fe] of
the RGB-a population, it is still difficult to understand why the MP population
(metal-poor, with [Fe/H]$\simeq$--1.7~dex) happens to have lower [$\alpha$/Fe] as
well.

Finally, although no direct He measurement for the RGB-a exists \citep[In][only
stars up to the intermediate population were analyzed]{dupree11}, it has been
suggested by various authors \citep{norris04,sollima05b,piotto05,renzini08} that
the most metal-rich population in $\omega$~Cen should possess an enhanced helium
content (ranging from Y=0.35 to Y=0.40, depending on the author). While this is
understandable from the stellar evolution and nucleosynthesis point of view
\citep{romano10}\footnote{In all the proposed self-enrichment scenarios, the
most metal-rich population should also be the youngest \citep[but see][where it
appears roughly coeval to the other populations]{sollima05b}, formed from gas
already enriched in helium. Thus, in these scenarios, the RGB-a/SGB-a population
should have at least the same helium abundance as the intermediate
populations.}, if true it would pose a problem when one attempts to determine
abundances from spectra using atmospheric models with a normal helium content.
In Paper~I, which dealt mainly with stars having [Fe/H]$<$--1.0~dex and
Y$<$0.30, we provided an approximated estimate of the impact of helium on
abundance calculations and found it irrelevant. When it comes to the SGB-a,
however, a deeper analysis is required.

In this paper we analyse a sample of high-resolution spectra of SGB-a stars,
selected from different photometry catalogues
\citep{p00,p03,sollima05a,villanova07} to represent the SGB-a component.  The
plan of the paper is the following: the data are presented in
Section~\ref{sec-data}; the details of the abundance analysis are described in
Section~\ref{sec-abo}; the abundance results are discussed in
Section~\ref{sec-res}; and in Section~\ref{sec-concl} we summarize our results
and draw our conclusions.

\section{Observations and data reduction}
\label{sec-data}

Observations were carried out betwen 27 and 29 April 2007, and the full dataset
and data treatment are decribed in detail by \citet{monaco10}. Among their data,
we selected only stars clearly belonging to the SGB-a \citep[as defined by][and
showed in Figure~\ref{fig_cmds}]{ferraro04} in all available photometries. The
final list of targets, together with some basic information (see also
Section~\ref{sec-param}) is reported  in Table~\ref{tab_logs}. The spectra were
observed with GIRAFFE@VLT, using the HR15n setup (6430--6810 \AA), because the
main goal of the original observations was the measurement of the lithium line at
6708~\AA. The resolution was R$\simeq$17\,000, the S/N varied between 25 and 50
(see Table \ref{tab_logs}). Most our target were observed in one single exposure
lasting 2 hours, while some of them were observed 9 times with a total exposure
time of 17.3 hours (see Table~\ref{tab_logs}).

\begin{table}
\caption{Equivalent widths and atomic data for the lines used in the classical
abundance analysis of the program stars. The complete version of the Table is
available at CDS. Here we show a few lines to illustrate its
contents.}             
\label{tab-ew}      
\centering          
\begin{tabular}{c c c c c c c c}     
\hline\hline 
Star & $\lambda$ & El & $\chi_{\rm{ex}}$ & $\log gf$ & EW & $\delta$EW & Q \\
 & (\AA) & & (eV) & (dex) & (m\AA) & (m\AA) & \\
\hline     	     
213129 & 6499.65 & CaI & 2.52 & -0.72 &  94.7 & 0.09 & 0.915 \\
213129 & 6572.78 & CaI & 0.00 & -4.10 &  44.0 & 0.19 & 1.291 \\
213129 & 6717.68 & CaI & 2.71 & -0.60 &  95.6 & 0.08 & 0.532 \\
213129 & 6469.19 & FeI & 4.83 & -0.77 &  62.0 & 0.20 & 1.533 \\
213129 & 6475.62 & FeI & 2.56 & -2.94 &  91.7 & 0.14 & 1.459 \\
\hline \hline                 
\end{tabular}
\end{table}

In short, the data were reduced with the 2.13 version of the GIRAFFE data
reduction pipeline\footnote{http://girbldrs.sourceforge.net/}. Seventeen fibres
were allocated to sky observations on each plate, and the average of the closest
fibers was subtracted from each target spectrum. After correcting for radial
velocity differences with the IRAF\footnote{http://iraf.noao.edu/. IRAF is
distributed by the National Optical Astonomy Ob\-ser\-va\-to\-ries, which is
operated by the association of Universities for Research in Astronomy, Inc.,
under contract with the National Science Foundation.} task {\em fxcor}, multiple
spectra of each target were averaged together. A few stars which significantly
deviated in their heliocentric velocity (derived with the IRAF task {\em
rvcorrect}) from the cluster average \citep[V$_r$=232.8 or 233.4 km/s, determined
by][respectively]{meylan95,p07} were rejected. The final number of spectra
surviving selection was 18.

\section{Abundance Analysis}                                                                   
\label{sec-abo}                                                                                

\subsection{Equivalent widths and atomic data}
\label{sec_ew}

\begin{figure}
\centering
\includegraphics[width=\columnwidth]{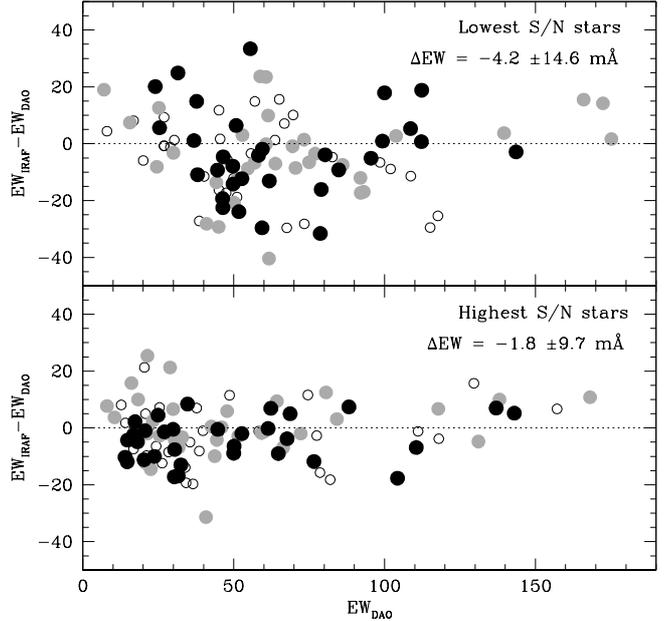}
\caption{Comparison between DAOSPEC and IRAF measurements for all elements. The
top panel shows the three highest S/N stars WFI~215700 (black points),
WFI~224921 (white points), and WFI~248814 (grey points). The bottom panel shows
the three lowest S/N stars WFI~214198 (black points), WFI~220401 (grey points),
and WFI~234254 (white points). Perfect agreement is marked with a dotted line in
both panels, along with the average differences and sigmas.}
\label{fig_ewhand}
\end{figure}

We selected the majority of our lines and their atomic data from the
VALD\footnote{http://www.astro.uu.se/$\sim$vald/} database \citep{vald}. The
NIST\footnote{http://physics.nist.gov/PhysRefData/ASD/index.html} atomic data
were employed for spectral syntesis of the Ba~II line at 6496~\AA, including
hyperfine structure (HFS) and isotope splitting. To identify reliable lines, the
unfiltered linelist of Paper~I was compared to the spectral range of the GIRAFFE
spectra, and only reliable, unblended, and relatively strong lines were
retained. DAOSPEC\footnote{Available at
http://www.bo.astro.it/$\simeq$pancino/projects/daospec.html,
http://www2.cadc-ccda.hia-iha.nrc-cnrc.gc.ca/community/STETSON/d aospec/}
\citep{daospec} was used to measure equivalent widths (EW) of all the chosen
lines. A first pass abundance analysis was performed (Section~\ref{sec_abo}):
lines that showed systematically higher errors and bad Q parameters
\citep[see][for details]{daospec,p10} and that simultaneously gave
systematically discrepant abundances were rejected. Finally, all lines that had
EW$<$15~m\AA, or EW$>$150~m\AA\   were not used to determine abundances. The
DAOSPEC EW measurements used for the abundance analysis are shown in
Table~\ref{tab-ew}, along with the formal error $\delta$EW and the quality
parameter Q for each line \citep[see][for details]{daospec}. 

\begin{figure}
\centering
\includegraphics[width=\columnwidth]{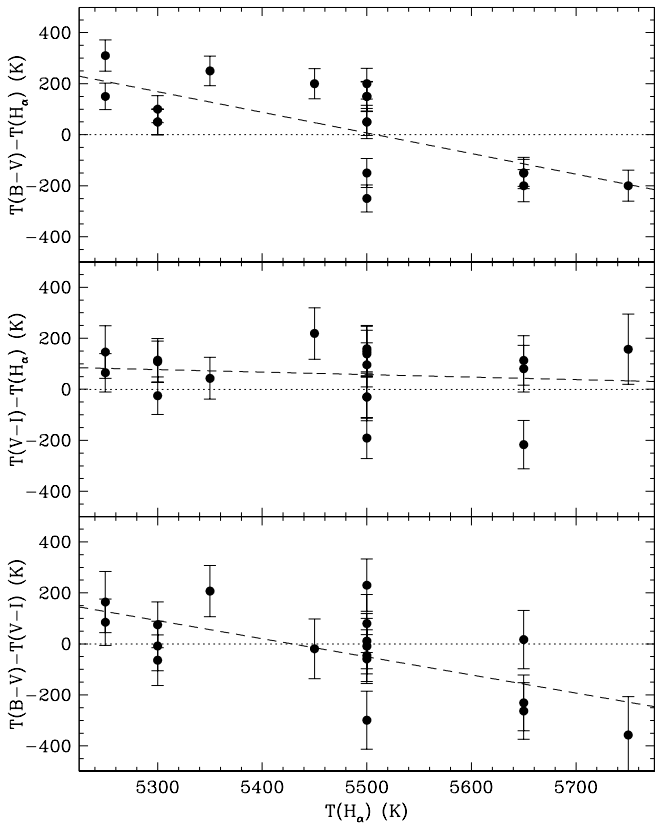}
\caption{Comparison between photometric temperatures based on the B$-$V and V$-$I colours
and the spectroscopic ones based on the H$_{\alpha}$ profile fitting (see text).}
\label{fig_temp}
\end{figure}

To check our DAOSPEC measurements, we re-measured the EW of six stars
(Figure~\ref{fig_ewhand}) with the IRAF task {\em splot}. The average EW
difference, computed on a total of 218 lines, was $-$1.8$\pm$9.7~m\AA\  for the
three highest S/N stars and $-$4.2$\pm$14.6~m\AA\  for the three lowest S/N
stars. All differences indicate that the IRAF measurements were slightly lower
than the DAOSPEC ones, but the two sets of EW are totally compatible within the
uncertainties.

\subsection{Atmospheric parameters}
\label{sec-param}

The HR15n setup of GIRAFFE (643--681~nm, approximately) contains few lines for
these subgiants, since it was mainly chosen to measure the Li abundance and to
have a precise estimate of T$_{eff}$ \citep{monaco10} using the H$_{\alpha}$ line
wings. Therefore, we could not perform a full spectroscopic analysis as in
Paper~I, and we adopted fixed atmospheric parameters for our stars. 

For temperatures, we used two independent estimates. The former is based on the
FORS photometry \citep{sollima05a}, where dereddened (B--V)$_0$ and (V--I)$_0$
colors were obtained from B, V and I magnitudes adopting E(B--V)=0.11
\citep{lub02} and E(V--I)/E(B--V)=1.30 \citep{dean78}, and are listed in
Table~\ref{tab_logs} along with the V magnitudes. The V--I color was converted
from the original V--I$_C$, based on the Cousins I magnitude, to the V--I$_J$,
based on the Johnsons I magnitude, with the relations by \citet{bessell79}.
Effective temperatures (hereafter T$_{eff}$) were then obtained with the
\citet{alonso99} calibration. The latter independent method is based on the
profile fitting of the H$_{\alpha}$ line wings, employing a modified version of
the BALMER\footnote{The original version provided by Kurucz can be found at
http://kurucz.harvard.edu/.} code, which uses the self-broadening theory by
\citet{barklem00} and the Stark broadening by \citet{stehle99}. A comparison
between the resulting T$_{eff}$ values is shown in Figure~\ref{fig_temp}. The
average difference between photometric (B--V) temperatures and spectroscopic
H$_{\alpha}$ ones is +34$\pm$174~K, while the one between the (V--I) and
H$_{\alpha}$ temperatures is 61$\pm$119~K. Given the good agreement, we
arbitrarily\footnote{Even if the trend is probably not significant, it can be
seen in Figure~\ref{fig_temp} that temperatures derived from the (B--V) colour
do not show a flat difference with temperatures determined with H$_{\alpha}$ or
(V--I). This could be due to a small residual colour term in the the B magnitude
calibration by \citet{sollima05a}.} decided to adopt the H$_{\alpha}$
temperatures for our analysis (Table~\ref{tab_logs})\footnote{The numbers in
Table~\ref{tab_logs} differ slightly from those in Table~2 by \citet{monaco10},
because here we used synthetic spectra built with atmosphere models of --1.0~dex
in metallicity, while there they showed values derived assuming --1.5~dex.},
with an estimated uncertainty of the order of $\pm$100~K.

\begin{table*}
\caption{Abundance ratios with random uncertainties (see text).}
\label{tab_abo}
\begin{center}
\begin{tabular}{c c c@{$\pm$}c c@{$\pm$}c c@{$\pm$}c c@{$\pm$}c c@{$\pm$}c c@{$\pm$}c c@{$\pm$}c} 
\hline\hline
\noalign{\smallskip}
ID$_{WFI}^a$ & ID$_{FORS}^b$ & \multicolumn{2}{c}{[Fe/H]} & \multicolumn{2}{c}{[Al/Fe]}
   & \multicolumn{2}{c}{[Ba/Fe]} & \multicolumn{2}{c}{[Ca/Fe]} 
   & \multicolumn{2}{c}{[Ni/Fe]} & \multicolumn{2}{c}{[Si/Fe]} 
   & \multicolumn{2}{c}{[Ti/Fe]} \\
& & \multicolumn{2}{c}{(dex)} & \multicolumn{2}{c}{(dex)} & \multicolumn{2}{c}{(dex)} 
   & \multicolumn{2}{c}{(dex)} & \multicolumn{2}{c}{(dex)}    
   & \multicolumn{2}{c}{(dex)} & \multicolumn{2}{c}{(dex)} \\
\noalign{\smallskip}
\hline
\noalign{\smallskip}
213129 & 33807 & --0.45 & 0.14 & +0.39 & 0.17 & 0.82 & ... &  +0.18 & 0.16 & --0.14 & 0.28 & +0.47 & ... &  +0.29 & 0.54 \\
213295 & 40145 & --0.59 & 0.08 & +0.08 & 0.08 & 0.88 & ... &  +0.34 & 0.15 & --0.14 & 0.19 & +0.54 & ... &  +0.38 & 0.33 \\ 
214198 & 40987 & --0.55 & 0.12 & +0.39 & 0.24 & 0.53 & ... &  +0.35 & 0.24 &  +0.32 & 0.14 & +0.81 & ... &  +0.75 & 0.40 \\ 
215315 & 65607 & --0.71 & 0.13 & +0.51 & 0.22 & 0.59 & ... &  +0.08 & 0.19 & --0.07 & 0.16 & +0.64 & ... &  +0.51 & 0.48 \\ 
215700 & 37198 & --0.96 & 0.07 & +0.46 & ...  & 1.12 & ... &  +0.36 & 0.12 &  +0.24 & 0.10 & +0.81 & ... &  +0.55 & 0.29 \\ 
215931 & 61617 & --0.79 & 0.12 & +0.28 & ...  & 0.60 & ... &  +0.10 & 0.23 & --0.18 & 0.21 & +0.72 & ... &  +0.37 & 0.44 \\ 
216031 & 34467 & --0.75 & 0.07 & +0.34 & 0.07 & 0.79 & ... &  +0.43 & 0.09 &  +0.10 & 0.09 & +0.84 & ... & --0.03 & 0.32 \\ 
218364 & 29483 & --0.52 & 0.11 & +0.04 & 0.26 & 0.95 & ... &  +0.21 & 0.17 & --0.10 & 0.13 & +0.78 & ... &  +0.36 & 0.46 \\ 
220401 & 42942 & --0.70 & 0.15 & +0.26 & 0.28 & 1.17 & ... &  +0.31 & 0.23 &  +0.10 & 0.25 & +0.88 & ... &  +0.50 & 0.55 \\ 
220947 & 29233 & --0.65 & 0.09 & +0.21 & 0.14 & 0.90 & ... &  +0.23 & 0.13 &  +0.24 & 0.12 & +0.33 & ... &  +0.13 & 0.35 \\ 
224701 & 63733 & --0.86 & 0.15 & +0.30 & 0.15 & ...  & ... &  +0.17 & 0.20 & --0.36 & 0.18 & +0.71 & ... &  +0.47 & 0.61 \\ 
224921 & 33448 & --0.89 & 0.12 & +0.36 & 0.13 & 1.18 & ... &  +0.38 & 0.13 & --0.11 & 0.12 & +0.64 & ... &  +0.46 & 0.47 \\ 
227902 & 30356 & --0.67 & 0.19 & +0.47 & 0.19 & ...  & ... &  +0.02 & 0.22 & --0.27 & 0.27 & +0.58 & ... &  +0.63 & 0.68 \\
234254 & 33737 & --0.76 & 0.14 & +0.26 & ...  & 0.51 & ... &  +0.24 & 0.20 &  +0.11 & 0.22 & +0.98 & ... &  +0.59 & 0.51 \\ 
235569 & 62919 & --0.79 & 0.18 &   ... & ...  & ...  & ... & --0.08 & 0.27 & --0.08 & 0.25 & +0.52 & ... &  +0.61 & 0.66 \\  
243327 & 37860 & --0.72 & 0.07 & +0.29 & 0.09 & 0.99 & ... &  +0.47 & 0.11 & --0.08 & 0.13 & +0.70 & ... &  +0.16 & 0.34 \\  
247798 & 37337 & --0.89 & 0.09 &   ... & ...  & 0.97 & ... &  +0.51 & 0.11 &  +0.17 & 0.11 & +0.55 & ... &  +0.57 & 0.35 \\  
248814 & 38329 & --0.81 & 0.08 & +0.48 & 0.19 & 1.03 & ... &  +0.37 & 0.10 & --0.03 & 0.13 & +0.61 & ... &  +0.38 & 0.37 \\  
\hline\hline	
\end{tabular}
\end{center}
\end{table*}

Gravities were derived with the adopted T$_{eff}$ values, the BC$_V$ from
\citet{alonso99} as mentioned above, and a distance modulus
(m--M)$_V$=14.04$\pm$0.11~mag \citep{bellazzini04}, by means of fundamental
relations:

\begin{equation}
\label{eq_logg}
\log g_*=0.4(M_V+BC_V)+4\log T_{eff,*}-12.61
\end{equation}

\noindent where the solar values where assumed in conformity with the IAU
recommendations \citep{iau}, i.e., $\log g_{\odot}=4.437$,
T$_{eff,\odot}$=5770~K and M$_{bol,\odot}$=4.75. A typical mass of
0.8~M$_{\odot}$ was assumed for the program stars \citep{vdb01}. The final
log$g$ values are listed in Table~\ref{tab_logs}, and we estimated an
uncertainty of $\pm$0.2~dex. 

Finally, we verified that the microturbulent velocity, v$_t$, is poorly
constrained by the limited number of Fe lines at our disposal, so we adopted
v$_t$=1.0~km~s$^{-1}$ for all our SGB-a targets \citep[see][as well]{marino08},
as was also done by \citet{monaco10}, and we allowed for a conservative
uncertainty of $\pm$0.3~km~s$^{-1}$.

\subsection{Abundance calculations}
\label{sec_abo}

For all chemical species except Ba (see Section \ref{sec-esse}), we computed
abundances with the help of the updated version of the original code by
\citet{spite67}. Our reference solar abundance was \citet{gre96}. We used the
new plane-parallel MARCS\footnote{http://marcs.astro.uu.se/} model atmospheres
with standard composition\footnote{This means [$\alpha$/Fe]=+0.4 for metal-poor
stars of [Fe/H]$<$--1.0 and reaching [$\alpha$/Fe]=0 at [Fe/H]=0, following
schematically the typical halo-disk behaviour of the Milky Way field
population.}. We chose the closest available global model metallicity (taking
into account $\alpha$-enhancement) to the $\omega$~Cen sub-populations, which
was $-$1.0~dex for all targets.

For all species we computed a 3\,$\sigma$-clipped average of abundances
resulting from each available line. For Fe and Ti, which had both neutral and
ionized lines, we computed the weighted (on the number of lines) average of the
two ionization stages to obtain [Fe/H] and [Ti/Fe]\footnote{The abundance of the
few Ti~I and II lines were so scattered that the results presented in
Section~\ref{sec-alfa} would not change significantly if we used a straight
average.}. We typically rejected lines that had EW$>$150~m\AA, where the
Gaussian approximation could fail, or EW$<$15~m\AA, since the relative error was
too high. For Ca, one line at 6462~\AA\  was larger than 150~m\AA, but we
checked that the DAOSPEC measurement were not too underestimated by visually
inspecting the spectrum and overlaying the DAOSPEC Gaussian fit on each target
star.

\begin{table*}
\caption{Uncertainties due to the choice of stellar parameters.}
\label{tab_err}
\begin{center}
\begin{tabular}{l c c c c c c c} 
\hline\hline
\noalign{\smallskip}
& \multicolumn{3}{c}{Star WFI~216031} & \multicolumn{3}{c}{Star WFI~227902} & Average \\
El & $\pm$100~K & $\pm$0.3~km~s$^{-1}$ & Total & 
$\pm$100~K & $\pm$0.3~km~s$^{-1}$ & Total & $\Delta$[El/Fe]  \\
& (dex) & (dex) & (dex) & (dex) & (dex) & (dex) & (dex) \\
\noalign{\smallskip}
\hline
$\delta[$Fe/H$]$  & $\pm$0.06 & $\pm$0.02 & $\pm$0.06 & $\pm$0.05 & $\pm$0.02 & $\pm$0.05 & $\pm$0.06 \\
$\delta[$Al/Fe$]$ & $\pm$0.05 & $\pm$0.12 & $\pm$0.13 & $\pm$0.05 & $\pm$0.13 & $\pm$0.14 & $\pm$0.14 \\
$\delta[$Ba/Fe$]$ & $\pm$0.09 & $\pm$0.06 & $\pm$0.11 & $\pm$0.37 & $\pm$0.37 & $\pm$0.52 & $\pm$0.32 \\
$\delta[$Ca/Fe$]$ & $\pm$0.10 & $\pm$0.08 & $\pm$0.13 & $\pm$0.09 & $\pm$0.05 & $\pm$0.10 & $\pm$0.12 \\
$\delta[$Ni/Fe$]$ & $\pm$0.09 & $\pm$0.09 & $\pm$0.12 & $\pm$0.09 & $\pm$0.03 & $\pm$0.09 & $\pm$0.11 \\
$\delta[$Si/Fe$]$ & $\pm$0.00 & $\pm$0.02 & $\pm$0.02 & $\pm$0.02 & $\pm$0.02 & $\pm$0.03 & $\pm$0.03 \\
$\delta[$Ti/Fe$]$ & $\pm$0.07 & $\pm$0.03 & $\pm$0.08 & $\pm$0.06 & $\pm$0.05 & $\pm$0.08 & $\pm$0.08 \\
\noalign{\smallskip}                                   
\hline\hline	
\end{tabular}
\end{center}
\end{table*}

\subsection{Abundance uncertainties}
\label{sec-err}

For those elements that had more than one line after $\sigma$-clipping, we
estimated the random (internal) uncertainty as $\sigma / \sqrt n$, as reported
in Table~\ref{tab_abo}. When only one line was available, we put an ellipsis in
Table~\ref{tab_abo}, and a rough estimate of the associated random uncertainty
is $\simeq$0.10~dex, according to the \citet{cayrel88} formula.

Another source of uncertainty is the global accuracy of the continuum
normalization. We used the r.m.s. of the residual spectrum calculated by
DAOSPEC after removing all the fitted spectral lines, which was on average
$\pm$4\% for our GIRAFFE spectra. According to Figure~2 by \citet{daospec},
this corresponds roughly to a constant error of $\pm$10~m\AA, which in turn
corresponds to approximately $\pm$0.10~dex in our abundances.

Finally, to estimate the uncertainty due to the choice of atmospheric
parameters, we cannot use the \citet{cayrel04} method, since we did not
determine our parameters with the classical spectroscopic optimization method.
In our case, T$_{eff}$ and v$_t$ are determined in a fully independent way,
while log$g$ is fixed by the choice of T$_{eff}$ (see Section~\ref{sec-param}).
Therefore, we re-computed our abundances by altering separately T$_{eff}$ (by
$\pm$100~K) and v$_t$ (by $\pm$0.3~dex) for one of the coolest (WFI~216031) and
of the warmest (WFI~227902) stars. For each of the $\pm$100~K models, we adopted
the appropriate value of log$g$ from Equation~\ref{eq_logg}. We computed
therefore the uncertainties due to T$_{eff}$ as the average of the abundance
variations for the +100~K and the --100~K models, and we did the same for v$_t$.
The two uncertainties were then summed in quadrature (since the two parameters
are determined in a fully independent way) to yield an uncertainty for each of
the two stars, and the average of the two stars was taken as our estimate of the
parameters choice impact on our abundance ratios, as listed in
Table~\ref{tab_err}. For those elements with two ionization stages (Ti and
Fe), we computed the abundance ratio exactly as described in
Section~\ref{sec_abo}, i.e., using a weighted average on the two ionization
stages.

\subsection{Literature comparison of star WFI~214198}
\label{sec-v07}

We found one star in common with \citet{villanova07}, which is star WFI~214198 in
the \citet{p00} WFI catalogue, corresponding to star 28448 in their Table~1. The
atmospheric parameters agree well, because they found T$_{eff}$=5400~K,
log$g$=4.1~dex, and v$_t$=1.0~km~s$^{-1}$. They used the \citet{gre98} solar
composition, which is practically identical to our \citet{gre96}. 

However, we found a significant difference in the [Fe/H] ratio, because they
measure [Fe/H]=--1.12$\pm$0.08($\pm$0.15--0.20)~dex, while we found
--0.55$\pm$0.16($\pm$0.06)~dex, where the uncertanties in parenthesis are
systematic, due to the uncertainty in stellar parameters. Concerning other
elements, we found: $\Delta$[Ca/Fe]=+0.20~dex, $\Delta$[Ti/Fe]=--0.45~dex,
$\Delta$[Ba/Fe]=0.46~dex, where all differences are computed by subtracting our
measurements from the \citet{villanova07} ones. If we compare the ratios with
respect to hydrogen, we find: $\Delta$[Ca/H]=--0.27~dex,
$\Delta$[Ti/H]=--0.92~dex, $\Delta$[Ba/H]=--0.01~dex. Therefore, given the
uncertainties of both studies, we can conclude that the calcium and barium ratios
are compatible with each other, while the titanium discrepancy is most probably
due to our uncertainties for this element being significantly higher than for
other elements. 

The discrepancy in the iron abundances requires instead a deeper discussion.
Given that we used the same atmospheric parameters, similar solar composition,
and atmospheric models with a similar thermal structure, we can only ascribe the
discrepancy to the different spectral quality, because the \citet{villanova07}
analysis is based on R$\simeq$6\,400 spectra (ours have R$\simeq$17\,000),
although with higher S/N$\simeq$100-150 (ours have S/N$\simeq$25--50). In
particular, for deriving their [Fe/H] ratios, they analyzed a short and blue
spectral region (4400--4425 \AA)\footnote{Other elements were measured from a
handful of lines in slightly different spectral regions, and that probably
explains why their abundances are less different from ours.}, where metal lines
blanketing is high, and highly uncertain \citep{kurucz92,munari05,bertone08}.
Therefore, the uncertainty in the continuum placement can become even more
problematic than usual\footnote{ \citet{villanova07} used full spectral synthesis
of the spectral region between 4400--4425 \AA\  to derive [Fe/H], thus the
``overblanketing'' predicted by theoretical log$gf$ (from the Kurucz linelist) of
tiny iron lines could in principle induce a continuum misplacement, causing an
underestimate of the [Fe/H] ratio, such as is evident in the comparison presented
here. A quantitative analysis of the ``overblanketing'' effect is out of the
scope of the present paper, but the effect goes in the right direction.},
especially at low resolution. In this respect, we recall here that a different
low resolution study of SGB-a stars, based on the infrared calcium triplet
\citep{sollima05b}, found an average [Fe/H]$\simeq$--0.6~dex. On the other hand,
the SGB-a (or branch D), clearly merges with the RGB-a in all published
photometries \citep{bedin04,villanova07,bellini10}, and all high-resolution
measurements of RGB-a stars provide abundances higher than [Fe/H]$>$--1.0~dex
\citep[see, e.g.][]{p02,p04,johnson10,marino11}. In conclusion, we are confident
that our measurements are correct, within the quoted uncertainties.

\section{Abundance results}
\label{sec-res}

\begin{figure}
\centering
\includegraphics[width=\columnwidth]{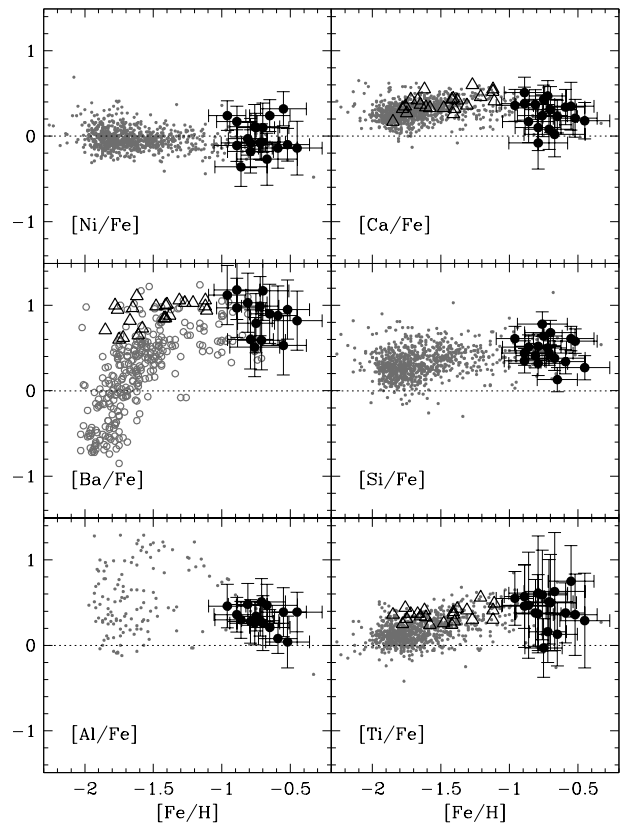}
\caption{Abundance ratios for all measured elements. In all panels, grey dots
are the measurements by \citet{johnson10}, grey empty circles by
\citet{marino11}, the empty triangles the ones by \citet{villanova07}, and the
black dots with errorbars our measurements.}
\label{fig_ratios}
\end{figure}

Among the several literature studies which provide abundance ratios, we selected
two to compare with our results, since they are the ones that contain the
largest samples of metal-rich ([Fe/H]$>$--1.0~dex) stars. The former is the
study of more than 800 red giants by \citet{johnson10}, which has similar
resolution to the GIRAFFE spectra presented here, and agrees well with most
previous studies (see their Figure~6) so it serves as a good comparison dataset.
The latter is the low resolution study by \citet{villanova07}, which studies
subgiant stars very similar to the ones analysed here (but see
Section~\ref{sec-v07}). In addition, we found a large sample of [Ba/Fe]
determinations of red giants at all metallicities in \citet{marino11}, based on
GIRAFFE spectra as well. These abundance ratios are plotted in
Figure~\ref{fig_ratios} along with our measurements, and are discussed in the
following Sections. While we corrected for the solar abundance reference values
differences, and we tried to check log$gf$ values when possible, it is clear
that residual zeropoint dfferences might be present in the comparison. 

\subsection{Iron-peak elements}
\label{sec-fepeak}

We could measure $\sim$10--15 Fe~I lines, depending on the star, and only one
Fe~II line, which gave discordant abundances and was discarded. We found an
average $<$[Fe/H]$>$=--0.73$\pm$0.14~dex. As discussed in Section~\ref{sec-v07},
a significantly lower iron abundance was found by \citet{villanova07}, based on
low-resolution spectra, but our measurements agree very well with past abundance
determinations of the RGB-a component
\citep{p02,origlia03,p03,p04,johnson10,marino11}, and with the SGB-a abundance
determination by \citet{sollima05b}. 

We could also measure 3--4 Ni lines, depending on the star, and we found an
average $<$[Ni/Fe]$>$=--0.02$\pm$0.19~dex, which agrees as well with the results
by \citet{johnson10}. A solar ratio of [Ni/Fe] is of course expected because Ni
is produced in the same site as Fe.

\subsection{Lithium}
\label{sec-li}

\begin{table}
\caption{Upper limits on Li abundances, for 2$\sigma_{ew}$. }
\label{litab}
\begin{center}
\begin{tabular}{ccccc}
\hline
 ID$_{WFI}$ & T$_{eff}$ & log\,$g$ & A(Li) & A(Li) \\
            &   K   & c.g.s. & 1D LTE & 3D NLTE \\          
\hline\hline
  213129  & 5300 &3.80 &$<1.31$&$<1.23$\\ 
  213295  & 5500 &3.90 &$<1.35$&$<1.30$ \\
  214198  & 5450 &4.00 &$<1.55$&$<1.49$ \\
  215315  & 5650 &3.90 &$<1.58$&$<1.56$ \\
  215700  & 5500 &4.00 &$<1.30$&$<1.25$ \\
  215931  & 5650 &4.00 &$<1.55$&$<1.52$ \\
  216031  & 5250 &3.70 &$<1.12$&$<1.04$ \\
  218364  & 5300 &3.70 &$<1.34$&$<1.27$ \\
  220401  & 5250 &3.90 &$<1.42$&$<1.32$ \\
  220947  & 5500 &3.80 &$<1.44$&$<1.40$ \\
  224701  & 5500 &3.90 &$<1.42$&$<1.38$ \\
  224921  & 5500 &3.90 &$<1.33$&$<1.28$ \\
  227902  & 5750 &4.10 &$<1.69$&$<1.67$ \\
  234254  & 5350 &3.90 &$<1.41$&$<1.34$ \\
  235569  & 5650 &4.00 &$<1.53$&$<1.49$ \\
  243327  & 5300 &3.70 &$<1.17$&$<1.10$ \\
  247798  & 5500 &4.00 &$<1.34$&$<1.29$ \\
  248814  & 5500 &4.00 &$<1.36$&$<1.31$ \\
\hline
\end{tabular}
\end{center}
\end{table}

We derived lithium upper limits from the 6708~\AA\  Li~I line (Table~\ref{litab})
of all the 18 SGB-a stars presented in this paper\footnote{The limits are derived
with a different method, assumed metallicity and parameters than in
\citet{monaco10}, and this explains the small differences in the two sets of
upper limits, for the 7 stars in common. The \citet{sbordone10}  fitting formulas
we used are based on 3D {\sf CO$^5$BOLD} models \citep{freytag02,freytag03} and
1D LHD models,  while \citet{monaco10} used ATLAS models. As can be seen, such
differences are unimportant.}. We used the fitting formulas of \citet{sbordone10}
for 1D LTE and 3D NLTE: the difference among the two is very small. The upper
limits correspond to the lithium abundance assuming an equivalent width of
$2\sigma_{EW}$, where $\sigma_{EW}$ was computed from the S/N ratio of the
spectrum derived from the \citet{cayrel88} formula. The choice of 2$\sigma$
implies that the abundance of each line has a probability of 0.0455 to be larger
than the upper limit, if noise prevented us to detect it. For this to happen 18
times the probability is $0.0455^{18} \simeq 7\times 10^{-25}$. 

Placing a firm constraint on the lithium content of SGB-a stars has some
importance when related to the elusive helium abundance of this population. As
will be discussed in detail in Section~\ref{sec_he}, the RGB-a/SGB-a population
should possess a high helium abundance \citep{norris04,piotto05,renzini08}. All
the H-burning processes, where He is produced, happen at temperatures where Li is
destroyed. Therefore, He-rich stars should have a very low lithium content.  

What is important here is that we have no detection \citep[as was the case
in][as well]{monaco10}, with upper limits in the range A(Li)=1.0 to 1.7,
depending on T$_{eff}$. In subgiants with these temperatures Li is expected to
be slightly depleted with respect to the Spite plateau 
\citep{ryan98,mucciarelli11}, but still detectable precisely in the above
abundance range, with a well defined ``Li-ridge'' \citep[see figure 4
of][]{ryan98}. Our upper limits suggest that the SGB-a Li abundance is below 
the standard ``Li-ridge''.  Thus the absence of any Li detection in our sample
of SGB-a stars provides indirect support to the notion that these stars are
indeed He-rich.

Deeper observations of the hottest stars of the sample would be desirable to see
if any Li is at all detectable. A measured Li abundance would provide a very
strong constraint on the amount of Li-free material to be mixed with Li-normal
material at the time of the star formation. In turn this would provide the
necessary He abundance of the He-rich material in order to obtain the total He
abundance implied by the CMD and abundance information of the main sequences.
This would place strong constraints on the stars responsible for the nuclear
processing. In the chemical evolution model of \citet{romano10}, He is provided
mostly  by massive AGB stars (4--5 M$_\odot$), and the He-rich stars are formed
from almost pure AGB ejecta. This material would certainly be Li-free, thus a
detection of Li in the He-rich sub-populations, even at a low level, could rule
out this scenario.

\subsection{$\alpha$-Elements}
\label{sec-alfa}

There were half a dozen measurable Ca lines in our spectra, complemented by one
single Si line and by 3--4 Ti~I and Ti~II lines, that we averaged together to
produce [Ti/Fe]. Calcium and titanium appear to agree with previous literature
estimates, being $<$[Ca/Fe]$>$=+0.26$\pm$0.16~dex, and
$<$[Ti/Fe]$>$=+0.45$\pm$0.12~dex. As discussed in Section \ref{sec-intro}, while
the 0.2--0.3 dex decrease in $\alpha$-enhancement initially found by \citet{p02}
and \citet{origlia03} is not confirmed by more recent studies quantitatively,
nevertheless a decrease of 0.1 dex approximately can be seen in all sufficiently
sampled studies \citep{norris95,smith04,p04,johnson10}, and is indeed present in
our data.

The lone Si line in our spectral range, at 6721.85~\AA, has a theoretically
computed log$gf$ \citep{kurucz73}, rather than measured in laboratory. In the
abundance analysis of the Sun performed by \citet{p10} with the same models, the
same code, and the same initial linelist and log$gf$ system, this line gave an
overestimated Si abundance by 0.19~dex. Therefore, the [Si/Fe] values of
Table~\ref{tab_abo} have been lowered by that amount in Figure~\ref{fig_ratios},
and they appear to agree with the \citet{johnson10} data, within the
uncertainties.

Our titanium ratios also follow the trend in the \citet{johnson10} data, which
have a tendency to rise with [Fe/H] as found also by other authors \citep[][to
name a few]{norris95,smith00}. The large uncertainties on our Ti
measurements are due to the low number of Ti~I and Ti~II lines ($\simeq$4--5,
depending on the star) available in the studied spectral range; these lines are
small ($\simeq$20--30~m$\AA$) and with large EW errors ($\simeq$10--20\%); the
VALD log$gf$ values are identical to the NIST ones, that are classified as D and
thus not very accurate; as a result the abundances of the 4--5 line are largely
inconsistent, and their scatter is the main reason for the the huge errobars
appearing in Figure~\ref{fig_ratios}. Therefore, even if a few stars appear to
have [Ti/Fe]$\simeq$0~dex, we consider this more likely a measurement problem
than an intrinsic property of these stars.

\subsection{Heavy elements}
\label{sec-esse}

We could measure barium by means of spectral synthesis of the 6496 \AA\  line,
which was generally around 100 m\AA. We used
MOOG\footnote{http://verdi.as.utexas.edu/moog.html} \citep{moog} for spectral
synthesis, in combination with the same models and atmospheric parameters used
for the EW abundance analysis\footnote{Two stars were re-analyzed with MOOG
(WFI~213129 and 215700), using the same atmospheric models, linelist and atomic
data, and solar reference abundances. We found an [Fe/H] abundance ratio of
--0.48~dex (0.03~dex lower than in Table~\ref{tab_abo}) for WFI~213129 and
$-$0.94 (0.02~dex higher) for WFI~215700, so we concluded that the two abundance
calculation codes give fully compatible results.}. We found
$<$[Ba/Fe]$>$=+0.87$\pm$0.23 dex for the SGB-a population. Our results  agree
with most past studies of RGB-a stars, which found a [Ba/Fe]$\sim$+1.0 dex for
stars with [Fe/H]$>$--1.0 dex \citep[see][to name a
few]{norris95,smith00,vanture02}, in continuity with the MInt populations lying
around --1.5$<$[Fe/H]$<$--1.0 dex. NLTE corrections for our [Ba/Fe] measurements
should be lower than 0.1~dex \citep{korotin11}. The only study of SGB-a stars
providing some barium abundance was that of \citet{villanova07}, which appears to
follows the general trend of red giants studies, having [Ba/Fe]$\sim$+1.0~dex,
although their [Fe/H] ratios are different from the ones derived here (see
Sections~\ref{sec-v07} and \ref{sec-fepeak}).

\subsection{Anti-correlations}
\label{sec-anti}

The only representative of the proton capture elements in our spectral range is
Al, for which we measured the 6696, 6698~\AA\  doublet. We did not apply NLTE
corrections to our abundances, because the used doublet should be relatively
free from NLTE effects \citep{gehren04,andrievsky08}. It is interesting to
recall here that both \citet{johnson10} and \citet{marino11} found a tendency,
for stars richer than [Fe/H]$\simeq$--1.0~dex, to exhibit no (anti-)correlation
among the usual elements (Na, Al, Mg, C, N, and O). Only one homogeneous group
of stars is present in the \citet{marino11} data, which appears roughly solar in
oxygen, but highly enriched in Na ([Na/Fe]$\simeq$+1~dex). The RGB-a appears
also enriched in Al ([Al/Fe]$\simeq$+0.4~dex) in the \citet{johnson10} data.
Here we find a homogeneous Al abundance among our 18 SGB-a stars, with
$<$[Al/Fe]$>$=+0.32$\pm$0.14~dex, well compatible with the result of
\citet{johnson10} and -- indirectly -- with \citet{marino11} as well, with a
spread comparable to the measurement errors. We stress that the importance of
this result lies not in the exact average value of [Al/Fe] found, but in the
fact that it is homogeneous among our SGB-a targets. Our result supports the
finding that no (anti)-correlation appears to be present among these metal-rich
stars.

\subsection{Helium overabundance}
\label{sec_he}

\begin{figure}
\centering
\includegraphics[width=\columnwidth]{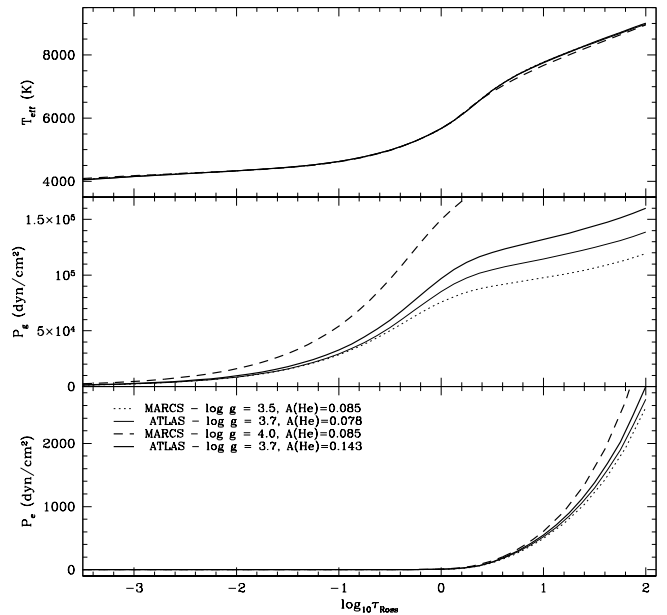}
\caption{Comparison of thermal and pressure structure of 5250~K MARCS models with
specially computed ATLAS~12 models. The top panel shows the temperature variation
with the logarithmic Rosseland opacity, the middle panel the gas pressure
variation, and the bottom panel the electron pressure variation. The MARCS models
have normal helium abundance and different gravities (log$g$=3.5~dex plotted as a
dotted line and log$g$=4.0~dex as a dashed line). The ATLAS~12 models have normal
helium (thin continuous line) and enhanced helium (thick continuous line) at a
fixed gravity of 3.7~dex. }
\label{fig_mods}
\end{figure}

It has been suggested by several authors \citep[starting
with][]{norris04,piotto05,renzini08}, that at least part of the MInt population
and the whole RGB-a population should be enriched in helium, with a typical
Y=0.35--0.40, depending on the author. Such a helium enriched atmosphere would be
different from the usual atmospheric models employed in abundance analysis
calculations \citep{bohm79}. A first order of magnitude evaluation of the impact
of helium can be obtained along the lines of Paper~I\footnote{In Paper~I, we used
the same formula, but we made a trivial error in the computation. We estimated
that, if the MInt star WFI~512115 was enriched in helium (Y$\simeq$0.35), we
would have underestimated its [Fe/H] by 0.08~dex because we used a Y$\simeq$0.25
atmospheric model. However, the correct value is much smaller: 0.03~dex. Such an
underestimate would be entirely within the measurement uncertainties.}, where we
used the \citet{gray} formula, and an exemplificative value of Y=0.35,
corresponding to a A(He)$\simeq$0.15 instead of 0.10:

$$\frac{\Delta g}{g} = \frac{4~\Delta A(\rm{He})}{1+4~A(\rm{He})}$$

We estimate that this would correspond to an increase -- for our targets -- from
log$g$$\simeq$4.0~dex to 4.2~dex. This would increase our abundances by
approximately 0.04~dex in [Fe/H], which is negligible, given the uncertainties
involved in the present analysis. 

To investigate the matter further, we computed helium enhanced atmosphere models
for three of our targets: one warm star (WFI~227902), one cool star (WFI~216031),
and one average star (WFI~220947). We kept the atmospheric parameters listed in
Table~\ref{tab_logs} fixed. We used the ATLAS~12 code\footnote{ATLAS~12 is the
only program publicly available to compute atmosphere models of arbitrary
chemical composition, so it was our only choice even if the rest of our abundance
analysis was made with MARCS models. In general differences between MARCS and
ATLAS models are negligible, once the appropriate mixing-length parameters are
chosen, taking into account the different formulation of the mixing-length theory
in the two codes \citep[see e.g.,][Appendix~A1]{bonifacio09}. }
\citep{kurucz05,castelli05} to compute models with a metallicity of -1.0~dex,
$\alpha$-enhanced, and the resulting models for star WFI~216031
(T$_{eff}$=5250~K, log$g$=3.7~dex) are compared with non-interpolated MARCS
models (log$g$=3.5 and 4.0~dex) in Figure~\ref{fig_mods}. As expected, the effect
of helium is similar to (but smaller than) the effect caused by an increase in
gravity: the thermal structure appears almost unaffected, while the pressure
structures are altered, with both gas and electron pressure increasing with
helium abundance and with gravity. 

\begin{table}
\caption{Abundance analysis with varying helium.}
\label{tab_he}
\begin{center}
\begin{tabular}{l c c c c} 
\hline\hline
\noalign{\smallskip}
WFI~216031 & ATLAS & $\Delta_{model}^a$ & ATLAS & $\Delta_{helium}^b$ \\
   & A(He)=0.085 &  & A(He)=0.143 & \\
\noalign{\smallskip}
\hline
$[$Fe/H$]$  & --0.73 &  +0.02 & --0.74 & --0.01 \\
$[$Al/Fe$]$ &  +0.33 & --0.01 &  +0.34 &  +0.01 \\
$[$Ca/Fe$]$ &  +0.44 & --0.01 &  +0.43 & --0.01 \\
$[$Ni/Fe$]$ &  +0.10 &   0.00 &  +0.12 &  +0.02 \\
$[$Si/Fe$]$ &  +0.85 &  +0.01 &  +0.87 &  +0.02 \\
$[$Ti/Fe$]$ & --0.02 &  +0.01 &  +0.01 & --0.01 \\
\noalign{\smallskip}                                   
\hline\hline	
\noalign{\smallskip}
WFI~220947 & ATLAS & $\Delta_{model}^a$ & ATLAS & $\Delta_{helium}^b$ \\
   & A(He)=0.085 &  & A(He)=0.143 & \\
\noalign{\smallskip}
\hline
$[$Fe/H$]$  & --0.64 & --0.01 & --0.65 & --0.01 \\
$[$Al/Fe$]$ &  +0.20 &  +0.01 &  +0.21 &  +0.01 \\
$[$Ca/Fe$]$ &  +0.23 &   0.00 &  +0.23 &  +0.00 \\
$[$Ni/Fe$]$ &  +0.23 &  +0.01 &  +0.25 &  +0.02 \\
$[$Si/Fe$]$ &  +0.33 &   0.00 &  +0.35 &  +0.02 \\
$[$Ti/Fe$]$ &  +0.16 &  +0.03 &  +0.13 & --0.03 \\
\noalign{\smallskip}
\hline\hline
WFI~227902 & ATLAS & $\Delta_{model}^a$ & ATLAS & $\Delta_{helium}^b$ \\
   & A(He)=0.085 &  & A(He)=0.143 & \\
\noalign{\smallskip}
\hline
$[$Fe/H$]$  & --0.66 & --0.01 & --0.65 &  +0.01 \\
$[$Al/Fe$]$ &  +0.47 &   0.00 &  +0.46 & --0.01 \\
$[$Ca/Fe$]$ &  +0.03 &  +0.01 & --0.36 & --0.01 \\
$[$Ni/Fe$]$ & --0.27 &   0.00 & --0.28 & --0.01 \\
$[$Si/Fe$]$ &  +0.58 &   0.00 &  +0.58 &  +0.00 \\
$[$Ti/Fe$]$ &  +0.65 &  +0.02 &  +0.66 &  +0.01 \\
\noalign{\smallskip}
\hline\hline
\multicolumn{5}{l}{$^a$Abundance ratio difference in the sense ATLAS~12 minus
MARCS.}\\                                         
\multicolumn{5}{l}{$^b$Abundance ratio difference of ATLAS~12 models in the sense
of en-}\\  
\multicolumn{5}{l}{hanced helium minus normal helium.}\\                                         
\end{tabular}
\end{center}
\end{table}

We then used the computed ATLAS~12 atmospheres to recompute our abundances for
the three chosen stars. We left the rest of the abundance analysis ingredients
untouched: we used the same atmospheric parameters, the same EWs and linelist,
the same atomic data, solar composition, and abundance calculation code. First we
used the helium-normal ATLAS~12 atmospheres to compare with the MARCS model
atmospheres analysis, then we compared the ATLAS~12 analysis with helium-normal
and enhanced atmospheric models. The results are presented in Table~\ref{tab_he},
where it can be appreciated that both the abundance ratio difference between the
MARCS and ATLAS~12 analysis, and between the ATLAS~12 analysis with normal and
enhanced helium, are negligible compared to the involved uncertainties, and
slightly smaller than our initial rough estimate of 0.04~dex in
[Fe/H]\footnote{Similarly small differences (smaller than $\pm$0.03~dex) were
found between He-normal and enhanced ATLAS~12 models using the Kurucz abundance
calculation code instead of the Spite one, when keeping all the remaining
ingredience of the calculation fixed.}. 

However, we did the above analysis by keeping the atmospheric parameters fixed. We
should have probably used a different set of temperatures and gravities. For
example, in the case of temperature, we used the H$_{\alpha}$ line wings profile
fit with a normal helium model atmosphere. If the atmosphere was enriched in
helium, we can again estimate an error of approximately 0.2~dex in the synthetic
spectrum gravity (an underestimate), which leads to an overestimate of the
temperature of less than 50~K for stars of the kind analysed here. According to our
calculations (see Table~\ref{tab_err}) this leads again to a negligible abundance
error of the same order of magnitude of those in Table~\ref{tab_he}. 

We can conclude that for the stars analysed here and in Paper~I, the effect of
using the wrong helium content in the atmospheric models produces errors that are
negligible compared to the uncertainties involved in the abundance analysis.

\section{Summary and Conclusions}
\label{sec-concl}

We analyzed spectra (R$\simeq$17\,000, S/N$\simeq$25--50) of 18 members of
$\omega$ Cen, lying on the SGB-a branch. {\em We found that
$<$[Fe/H]$>$=--0.72$\pm$0.14 dex, similarly to all past high-resolution studies
of the RGB-a population} \citep{p02,p03,johnson10,marino11}. The RGB-a is
clearly the bright-end continuation of the SGB-a according to all recent
high-quality photometries \citep{ferraro04,bedin04,villanova07,bellini10}. We
find some disagreement only with the low-resolution spectroscopic study by
\citet{villanova07}, and we ascribe that to the lower resolution of their
spectra, which cover a bluer range where line blanketing makes the continuum
positioning quite difficult, and overblanketing issues might lead to abundance
underestimates \citep{kurucz92,munari05,bertone08}. A similar low resolution
study \citep{sollima05b}, based on the infrared calcium triplet, gives an SGB-a
abundance in good agreement with our estimate.

{\em Abundance ratios of $\alpha$-elements were computed, finding
$<$[$\alpha$/Fe]$>$=+0.40$\pm$0.16 dex. This leaves little room for a significant
decrease of the $\alpha$-enhancement ($>$0.2~dex) of the RGB-a/SGB-a population
with respect to the more metal-poor ones.} A decrease of
$\Delta$[$\alpha$/Fe]$\simeq$0.2 and 0.3 dex was found by \citet{p02} and
\citet{origlia03} respectively, but according to \citet{p04} and
\citet{johnson10}, this decrease should be lower, of 0.1--0.15 dex at most. While
type Ia supernovae were invoked in the past to explain this decrease, a more
complex situation appears from these larger samples of stars, where the
$\alpha$-enhancement increases slowly with [Fe/H], reaches its maximum in the
MInt population, and then slightly decreases again in the RGB-a population. A
straightforward interpretation of this behaviour probably requires detailed
chemical evolution modelling. While the metal-rich end could still be explained
by type Ia supernovae intervention, the lower [$\alpha$/Fe] of the metal-poor
stars around [Fe/H]=--1.7~dex requires a variation in the star formation rate
during the chemical enrichment history of $\omega$~Cen.

We could also measure Al which -- together with C, N, O, Na, and Mg -- is one of
the elements that (anti)-correlate in nearly all GC studied up to now
\citep[see][and references therein]{gratton04,carretta09}. It is
interesting to note that {\em we found a fairly homogeneous Al abundance among
our SGB-a targets, in good agreement with what found by \citet{johnson10} and --
indirectly -- by \citet{marino11} with their O and Na measurements.} We
measured $<$[Al/Fe]$>$=+0.32$\pm$0.14 dex, where the spread is compatible with
measurement errors only. The importance of this result does not lie in the exact
value of the average [Al/Fe], but in the fact that it is homogenous among our 18
SGB-a stars. The field populations of dwarf galaxies and of the Milky Way are
substantially free from (anti)-correlations \citep{martell10}, even when they
are found among their GC \citep{letarte06,mucciarelli09}. Similarly, M 54 shows
a clear Na-O anti-correlation while the field population of the Sagittarius
dwarf galaxy is free of it \citep[see
e.g.,][]{monaco05,sbordone07,carretta10}\footnote{The case of Terzan~5
\citep{ferraro09,origlia11} has no similarity to what presented here, because
none of its two populations shows any anti-correlation and therefore this object
must have had an entirely different chemical evolution.}. 

In the popular scenario of the formation of $\omega$~Cen as a disrupted dwarf
galaxy \citep[see, e.g.,][and references therein]{bekki03}, it can be speculated
that, if any of the sub-populations of $\omega$ Cen is free from
(anti)-correlations, that population is a likely candidate for its putative
parent galaxy field population. This was discussed for the case of the VMP (very
metal-poor) population defined in Paper I, and was also discussed by \citet{p03}
and \citet{carretta10}. We have then two candidate populations for the parent
galaxy field: the VMP and the RGB-a/SGB-a. However, (anti)-correlations are not
completely ruled out in the case of the VMP \citep[see Paper
I;][]{johnson10,marino11}, and they could simply be of a lesser extent. 
Moreover, the metallicity of the Sagittarius field population around M 54 is
metal-rich \citep[about --0.5 dex,][]{carretta10}, analogously to the
SGB-a/RGB-a population in $\omega$ Cen, so the RGB-a/SGB-a should be the best
candidate. It must be noted, however, that the analogy with Sagittarius and
other dwarf galaxies breaks when the $\alpha$-elements are considered, being
higher in $\omega$~Cen than in any other dwarf. Also, the high [Na/Fe] of the
RGB-a/SGB-a population and its narrow metallicity range are not easily explained
in this scenario. Thus more work is required to assess if $\omega$~Cen is really
the remnant of an accreted and disrupted dwarf galaxy, and to find its elusive
relics in the field population of the Galaxy \citep[see,
e.g.,][]{meza05,dacosta08,sollima09}.

Finally, it has been speculated \citep[see, e.g.,][]{norris04,piotto05,renzini08}
that the RGB-a population should be rich in helium (with Y$\simeq$0.35--0.40),
and therefore we tested whether an usual abundance analysis, based on model
atmospheres with a normal (Y$\simeq$0.25) content, could significantly affect the
resulting abundance ratios. We calculated helium-enhanced models with ATLAS~12
and recomputed our abundances, and the results is that {\em the effect of the
helium content of the model atmospheres has a negligible impact on the resulting
abundance ratios for stars of the type studied here and in Paper~I.} We also
determined upper limits to the lithium content and found no 6708~\AA\  Li~I line
detection in our 18 stars. {\em This suggests that lithium in SGB-a stars is less
abundant than what is typical \citep{ryan98}, and therefore lends support to the
notion that these stars have enhanced helium content, as discussed in detail in
Section~\ref{sec-li}.}

\begin{acknowledgements} 

We warmly thank M. Bellazzini and F. R. Ferraro for their advice. EP would like
to acknowledge the hospitality of the Universidad de Concepci\'on, Chile, where
part of this work was carried out. We also thank an anonymous referee for
her/his work on this paper.

\end{acknowledgements}

\end{document}